\documentclass[twocolumn,amsmath,amssymb,prl,10pt,aps,floatfix]{revtex4}
\usepackage{graphicx}
\usepackage{dcolumn}
\usepackage{bm}
\usepackage{color}

\usepackage{dsfont}
\usepackage[T1]{fontenc}
\allowdisplaybreaks

\begin{document}

\title{Unique Spin Vortices in Quantum Dots with Spin-orbit Couplings}
\author{Wenchen Luo}
\email{wenchen.luo@umanitoba.ca}
\author{Amin Naseri}
\email{naserija@umanitoba.ca}
\author{Jesko Sirker}
\email{sirker@physics.umanitoba.ca}
\author{Tapash Chakraborty}
\email{Tapash.Chakraborty@umanitoba.ca}
\affiliation{Department of Physics and Astronomy, University of Manitoba, Winnipeg,
Canada R3T 2N2}
\date{\today}

\begin{abstract}
Spin textures of one or two electrons in a quantum dot with Rashba or
Dresselhaus spin-orbit couplings reveal several intriguing
properties. We show that even at the single-electron level spin
vortices with different topological charges exist. These topological
textures appear in the {\it ground state} of the dots. The textures
are stabilized by time-reversal symmetry breaking and are robust
against the eccentricity of the dot. The phenomenon persists for the
interacting two-electron dot in the presence of a magnetic field.
\end{abstract}

\maketitle


A variety of topological states have recently been observed in
condensed matter physics. These novel states of matter are a direct
consequence of spin-orbit coupling (SOC)
\cite{TP01,TP02} with topological insulators (TIs) being one of the most prominent
examples \cite{TP03,TP04}. The SOC also plays an important role in
tailoring topological superconductors (TSs) where the elusive Majorana
fermions might be present \cite{Mj01,Mj02,Mj03}. Both TIs and TSs
display a topologically non-trivial structure in momentum space. SOC
can, however, also lead to topological charges in real space. The
Dzyaloshinskii-Moriya interaction \cite{DM01,DM02}---microscopically
based on the SOC---can, for example, give rise to spin skyrmions in
helical magnets
\cite{sky01,sky02} and pseudospin skyrmions in bilayer graphene
\cite{bilayerg}. Synthetic spin-orbit couplings can also be
engineered in cold atomic gases and skyrmion-like spin textures have
been observed \cite{STB01,STB05}.

Quantum dots (QDs) are of practical and fundamental interest and
provide an excellent platform to control the spin and charge of a
single electron \cite{QD00,QD01,QD02,QD03,QD04,QD05,QD06,QD07}.
Extensive studies on QDs with SOCs have been reported in recent years
\cite{QDSOC01,QDSOC02,QDSOC03,QDSOC04,QDSOC05,QDSOC06,QDSOC07,QDSOC08,QDSOC09, QDSOC10,
QDSOC11,siranush,QDSOC12,QDSOC13,QDSOC14,QDSOC15,QDSOC16}. Furthermore, a Berry connection
\cite{Berr} in momentum space induced by the SOC has been studied \cite{QDSOC12,QDSOC13}.

In this letter, we investigate the spin textures associated with the
electron density profiles in isotropic and elliptical QDs. We show that
in the presence of SOC the in-plane spin texture of a single electron
is a spin vortex. The QD is consequently turned into an artificial
atom \cite{maksym} with topological features. Spin vortices often
emerge in many-spin systems forming either a crystalline arrangement
or vortex/anti-vortex pairs \cite{vor01,vor02}. For instance, in
quantum Hall systems the skyrmion is a single-particle excitation in
low Landau levels and the in-plane spin texture is similar to the one
we find in a QD with SOC. The skyrmion excitations in the former case
are, however, induced by Coulomb interactions \cite{Ezawa}. In
contrast, we show here that in a QD a single vortex can exist in the
ground-state of a non-interacting quantum system.

We focus on the physics of the two-dimensional (2D) surface where the
QD is constructed \cite{QD01}. We consider both the Rashba and the
linear Dresselhaus SOCs which arise in materials with broken inversion
symmetry. The strength of the Rashba SOC can be controlled by a gate
electric field
\cite{Rash01,Rash02,Rash03,Rash04,Rash05}. Moreover, the ratio of the
Rashba SOC to the Dresselhaus SOC can be tuned over a wide range, for
instance in InAs QDs, by applying an in-plane magnetic field
\cite{Rash05}. We will show that this leads to a system where
the topological charge can be dynamically controlled by external
electromagnetic fields making spin vortices in QDs possible candidates
for future spintronics and quantum information applications.

The SOCs can be theoretically considered as effective
momentum-dependent magnetic fields \cite{SOC01,SOC02,SOC03,SOC04}. In
the absence of a confinement and an external magnetic field, the
momentum is conserved and the SOC in the Hamiltonian becomes a
momentum-dependent operator with a good quantum number (e.g., the
helicity operator for Rashba SOC). On the other hand, the spin state
is momentum-independent if both Rashba and Dresselhaus couplings have
equal strength and there is no Zeeman coupling, leading to a
persistent spin helix \cite{shx01,shx02,shx03}. This particular spin
state persists in the presence of a confinement potential and can be
obtained by exactly solving the Hamiltonian which is equivalent to a
quantum Rabi model \cite{supplement}. If the spin is not a good
quantum number then it is instructive to study the spin field in a
given single-particle wavefunction $\Psi(\mathbf{r})$ of the dot
\begin{eqnarray}
	\sigma_{i}(\mathbf{r})
	=
	\Psi^{\dag}(\mathbf{r})
	\sigma_{i}
	\Psi(\mathbf{r}),
	\label{eq_SpDen01}
\end{eqnarray}
where $\sigma_{i}$ for $i=x,y,z$ are Pauli matrices. An in-plane
vector field $\boldsymbol{\sigma}(\mathbf{r}) = \left(
\sigma_{x}(\mathbf{r}), \sigma_{y}(\mathbf{r}) \right)$ reveals how
the spin in real space is locally affected by the effective magnetic
field. In the following, we demonstrate that generic SOCs compel the
spin field to rotate around the center of the QD and to develop into a
spin vortex.


The Hamiltonian of an electron with effective mass $m^{\ast }$ and charge $-e
$ in a quantum dot with SOCs is given by%
\begin{equation}
H=\frac{\left( \mathbf{p+}e\mathbf{A}\right) ^{2}}{2m^{\ast }}+\frac{m^{\ast
}}{2}\left( \omega _{x}^{2}x^{2}+\omega _{y}^{2}y^{2}\right) +\frac{\Delta
\sigma _{z}}{2}+H_{SOC},  \label{hamiltonian}
\end{equation}%
where the vector potential is chosen in the symmetric gauge $\mathbf{A}=%
\frac{1}{2}B\left( -y,x,0\right) $ with the magnetic field $B$. The
confinement is anisotropic with the frequencies in two directions,
$\omega _{x}$ and $\omega _{y}$, and $\Delta $ is the Zeeman coupling. We
consider both the Rashba SOC, $H_R$, and the Dresselhaus SOC, $H_D$, with
\begin{eqnarray}
H_{R} &=&g_{1}\left( \sigma _{x}P_{y}-\sigma _{y}P_{x}\right) , \\
H_{D} &=&g_{2}\left( \sigma _{y}P_{y}-\sigma _{x}P_{x}\right) ,
\end{eqnarray}%
and $H_{SOC}=H_{R}+H_{D}$. $P_{i}=p_{i}+eA_{i}$ is the kinetic
momentum, and $g_{1,2}$ determine the strength of each SOC. We note
that Rashba and Dresselhaus terms have different rotational symmetry
generators, $H_R$ commutes with $L_{z} + \hbar \sigma_{z}/2$ while $H_D$ 
commutes with $L_{z} -\hbar \sigma_{z}/2$, where
$L_{z}$ is the $z$-component of the angular momentum operator. In the
following, we will show that this difference is responsible for the
different topological charges associated with the spin vortex of the
dot.

It is also useful to introduce a renormalized set of frequencies
$\Omega _{i}=%
\sqrt{\omega _{i}^{2}+\omega _{c}^{2}/4}$ with the cyclotron frequency $%
\omega _{c}=eB/m^{\ast }$. The natural length scales in $x$ and $y$ directions are
$\ell _{i}=\sqrt{\hbar /(m^{\ast }\Omega _{i})}$ while the confinement
lengths are defined as $R_{i}=\sqrt{%
\hbar /(m^{\ast }\omega _{i})}$. In the numerical calculations presented in the
following the eigenvectors of $H_{0}=\frac{\mathbf{p}^{2}}{ 2m^{\ast}}+\frac{m^{\ast }}{2}
\left( \Omega _{x}^{2}x^{2}+\Omega_{y}^{2}y^{2}\right) +\frac{\Delta }{2}\sigma _{z}$, 
which is a two-dimensional harmonic oscillator, are used as a basis set.

\begin{figure}[htp]
\includegraphics[width=0.99\columnwidth]{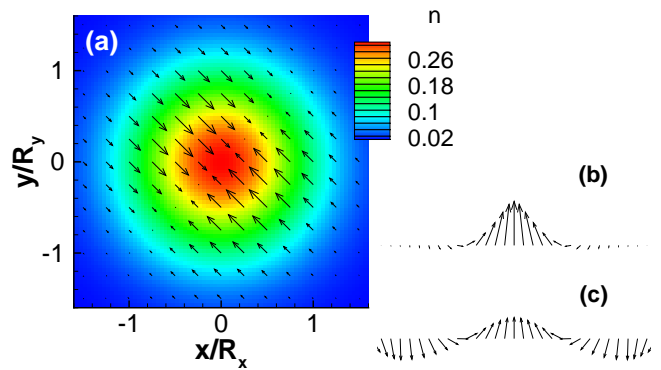}
\caption{(Color online) Numerical results for a single-electron QD
with $R_x=R_y=35$nm, $B=0.1$T, and equal SOCs $\hbar g_1=\hbar g_2=20$
nm$\cdot$ meV. (a) Electron density (contours) and in-plane spin fields (arrows), (b)
$\sigma_z (\mathbf{r})$ along $x=-y$, and (c) the normalized
$\tilde{\sigma}_z (\mathbf{r})=\sigma_z
(\mathbf{r})/\sqrt{\boldsymbol{\sigma} (\mathbf{r})^2+\sigma_z
(\mathbf{r})^2}$ along $x=-y$.}
\label{fig1}
\end{figure}

No analytical solution is known for the generic Hamiltonian in
Eq.~\eqref{hamiltonian} due to its complexity \cite{ExSOC}. We can,
however, analytically investigate the special case of an isotropic dot
($\Omega_{x,y}=\Omega$, $\ell_{x,y}=\ell$) without a magnetic field
and with equal SOCs, $g_{1,2}=g$. The Hamiltonian \eqref{hamiltonian}
is then equivalent to a two-component quantum Rabi model which has
been extensively studied in quantum optics \cite{supplement}. The
ground states in this case are a degenerate Kramers pair due to time
reversal symmetry,
\begin{equation}
\left\vert GS\right\rangle _{\pm }=\frac{1}{\sqrt{2}} e^{\pm i\sqrt{2}m^* 
\left( y-x\right) g/\hbar }\left(
\begin{array}{c}
\pm e^{-i \pi/4} \\
1
\end{array}%
\right) \left\vert 0,0\right\rangle
\end{equation}%
where $\left\vert 0,0\right\rangle$ is the ground state of the
two-dimensional quantum oscillator $H_0$. A very weak magnetic field
will lift the degeneracy of the Kramers pair, and the unique ground
state is then given by $\left\vert GS\right\rangle=(\left\vert
GS\right\rangle_+ + \textrm{sgn}(\Delta)
\left\vert GS\right\rangle_-)/\sqrt{2}$ which minimizes the energy \cite{supplement}.
The spin fields are consequently well defined. We note some features
of the spin field: (i) There is a mirror symmetry about the line
$x=\pm y$.  (ii) $ \sigma_x (\mathbf{r})+ \sigma_y (\mathbf{r}) =0$,
and $\sigma_x (\mathbf{r})=\sigma_y (\mathbf{r}) =0$ along the line $x=y$.
(iii) $ \sigma_z (\mathbf{r})=-\frac{\textrm{sgn}(\Delta)}{\pi\ell^2}
e^{-2x^2/\ell_x} \cos \left( 4\sqrt{2}m^* xg/ \hbar \right)$ along the
line $x=-y$, i.e., $\sigma_z (\mathbf{r})$ is a spiral. Its period is
related to the effective mass and the strength of the SOCs. We find
that the exact solution perfectly agrees with the exact diagonalization
results shown in Fig. \ref{fig1}.
Similar results are found for the case $g_1=-g_2$. For large magnetic
fields the exact solution for the case without field is no longer a
good starting point and the spin texture starts to rotate
\cite{supplement}.

\begin{figure}[htp]
\includegraphics[width=0.99\columnwidth]{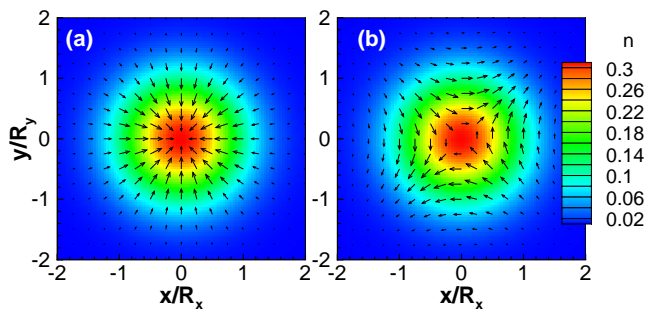}
\caption{(Color online) Single-electron QD with $R_x=R_y=15$nm, $B=0.1$T ($\Delta<0$), and (a) Rashba
SOC  $\hbar g_1=40$ nm$\cdot$ meV only, and
(b) Dresselhaus SOC $\hbar g_2=20$ nm$\cdot$ meV only.}
\label{fig2}
\end{figure}

Next, we study the case of an isotropic dot in a weak magnetic field
with generic strengths of the SOCs $g_1$ and $g_2$ based on a standard
perturbative calculation. We find that the in-plane spin fields up to
first order in $g_{1,2}$ are given by
\begin{eqnarray}
	\sigma_{x}(\mathbf{r})
	&=&
	\xi(r)	(r/\ell)
	\left(
	\bar{g}_{2}
	\sin{\theta}
	-
	\bar{g}_{1}
	\cos{\theta}
	\right),
	\\
	\sigma_{y}(\mathbf{r})
	&=&
	\xi(r) (r/\ell)
	\left(
	\bar{g}_{2}
	\cos{\theta}
	-
	\bar{g}_{1}
	\sin{\theta}
	\right),
\end{eqnarray}
and $\sigma_z(\mathbf{r})=\xi(r)/2$ with $\xi
(r)=2e^{-r^{2}/\ell^{2}}/\pi\ell^2 $, $\theta$ is the polar angle in
coordinate space, and the new parameters are
\begin{eqnarray}
	\bar{g}_{1,2}
	=
	\frac{\hbar g_{1,2}}{\ell}
	\frac{1 \pm \omega_{c} / (2\Omega) }{ \hbar (\Omega \pm \omega_{c}/2)- \Delta},
\end{eqnarray}
where we have assumed $\Delta<0$. The in-plane spin field
$\sigma(\mathbf{r})$ winds once around the origin and acquires a
topological charge $q=\pm 1$ when $\bar{g}_1 \neq \bar{g}_2$.  If $\bar{g}_{1} =
\bar{g}_{2}$, no vortex appears in agreement with the exact
solution discussed earlier. If ${g}_{1} = 0$ or ${g}_{2}=0$,
$\sigma(\mathbf{r})$ obtained perturbatively qualitatively agrees with
the numerical solutions shown in Fig.~\ref{fig2}, and the vortices even exist in
a strong magnetic field beyond the perturbation calculations.
We stress that the two vortex configurations are stable and
representative for the regime $g_1 \gg g_2$ and $g_2 \gg g_1$,
respectively \cite{supplement}. We further note that under $B\to -B$
the spin field changes direction, $\sigma(\mathbf{r})\to
-\sigma(\mathbf{r})$, leaving the topological charge invariant though.

Next, we analyze the rotational symmetry of the two types of SOCs in
order to characterize the sign of the winding number. First, we
consider the spin field of a dot when only the Rashba SOC is
present. The spin field is then invariant under the rotation matrix
\begin{eqnarray}
	U_{R}(\vartheta)
	=
	\begin{pmatrix}
      		\cos{\vartheta}&  \sin{\vartheta}  \\
      		-\sin{\vartheta}&   \cos{\vartheta}
	\end{pmatrix} ,
\end{eqnarray}
for $\vartheta \in [ 0, 2 \pi]$, which is rooted in the rotational
symmetry of a Rashba dot under the operator $L_{z}+\hbar
\sigma_{z}/2$. Therefore, the in-plane spin rotates clockwise by
$2\pi$ if we move around the center of the dot in a clockwise
direction, and hence, its winding number is $q=+1$. On the other hand,
the in-plane spin field of a dot with only Dresselhaus SOC being
present, is invariant under the action of
$U_{D}(\vartheta)=U_{R}(-\vartheta)$.  Along the same line of
reasoning, the in-plane spin field then rotates anticlockwise by
$2\pi$ if we move around the center in a clockwise direction.
Dresselhaus SOC thus leads to a winding number $q=-1$. In the absence
of an external magnetic field $B$, Kramers degeneracy may cancel the
spin textures, since there is a global $\pi$ phase difference between
the pair. Hence, the vortices should be stabilized by breaking of
time-reversal symmetry.

\begin{figure}[htp]
\includegraphics[width=0.7\columnwidth]{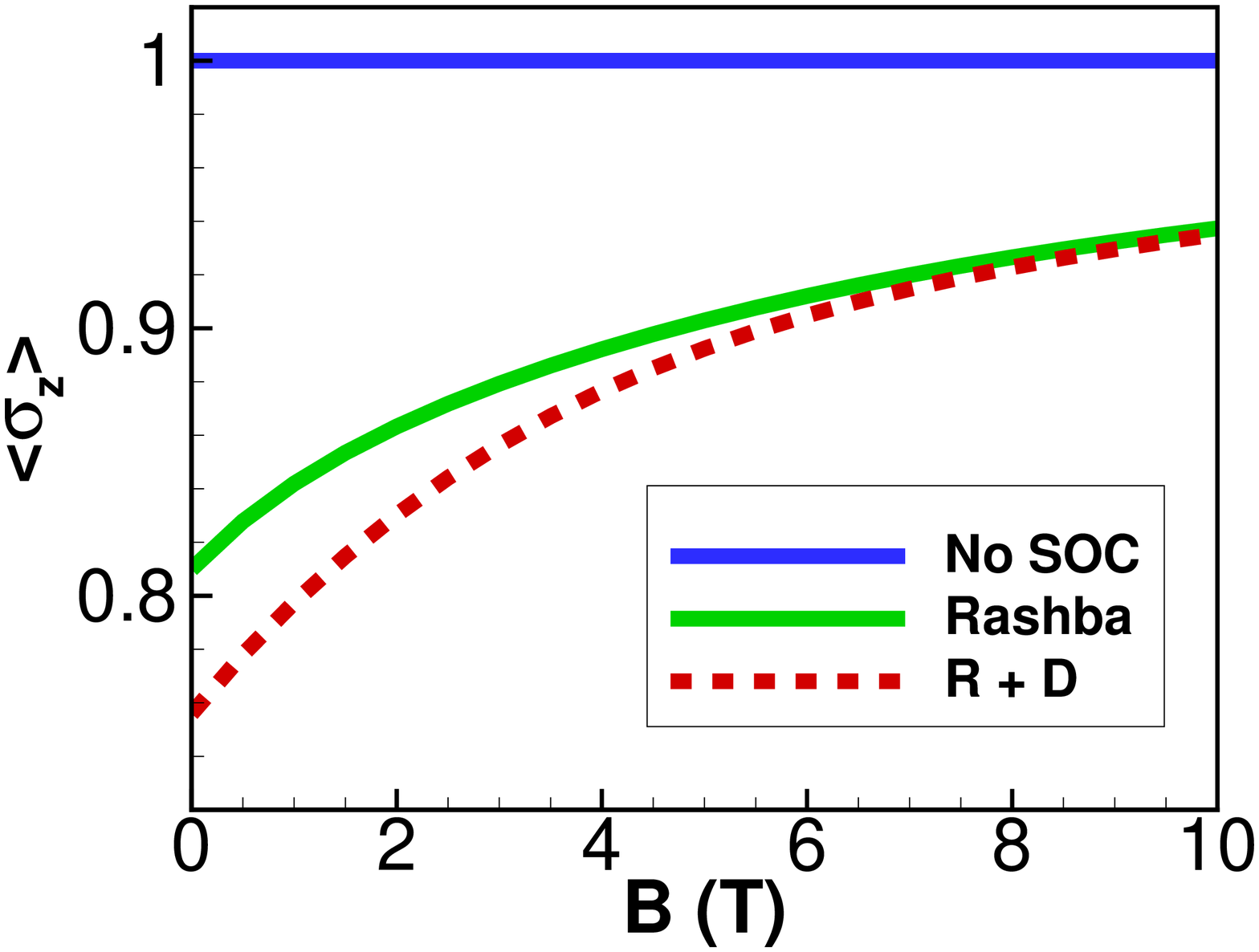}
\caption{(Color online) The total $\left\langle \sigma _{z}\right\rangle $
in a single-electron dot ($R_x=R_y=15$nm) without SOC, with Rashba SOC
only ($\hbar g_1=40$ nm$\cdot$ meV), and with both Rashba and
Dresselhaus SOCs ($\hbar g_1=40$ nm$\cdot$ meV, $\hbar g_2=20$
nm$\cdot$ meV).}
\label{fig3}
\end{figure}

In summary, we find for the single-electron dot with $g_{1}=\pm g_{2}$
and without or in a very weak magnetic field, that the in-plane spin
field does not form a vortex. There is, however, a spiral in $\sigma
_{z}\left(\mathbf{r}\right) $ along the line $x=\mp y$. For dominant
Rashba or Dresselhaus SOC, on the other hand, the exact
diagonalization results clearly show the formation of spin
vortices. Rashba SOC induces a vortex with topological charge $q=+1$
while the Dresselhaus SOC induces a vortex with $ q=-1$. These
topological charges associated with the spin textures are stabilized
by time-reversal symmetry breaking and are robust against the
ellipticity of the dot
\cite{supplement}. If the dot is strained, the topological features
are not changed, since the spin textures originate from the SOCs of
the material. The total $\langle\sigma_{z}\rangle$ in the presence of
SOC is no longer constant as a function of the applied magnetic field
and becomes more and more polarized with increasing magnetic field. In
Fig.~\ref{fig3} we compare $\left\langle \sigma _{z}\right\rangle $
for different cases.
The distinct behavior of $\left\langle \sigma _{z}\right\rangle $ when
SOCs are present might be observable experimentally via magnetometry
or optically pumped NMR measurements \cite{sean,private}.


If there is more than one electron confined in the dot, we need to
also consider the Coulomb interaction. The Hamiltonian of the
interaction is given by $ H_{C}=V\left( n_{1},n_{2},n_{3},n_{4}\right)
c_{n_{1}}^{\dag }c_{n_{2}}^{\dag }c_{n_{3}}c_{n_{4}}, $ where $c$ is
the electron annihilation operator and $n_{i}=\left(
n_{ix},n_{iy},n_{s}\right) $ is an index combining the quantum numbers
of the two-dimensional oscillator in $x,y$ direction with the spin
index. The interaction matrix elements are given in the
Suppl.~Mat. \cite{supplement}. The full Hamiltonian with interaction
is then $H_{I}=H+H_{C}$ with $H$ as given in
Eq.~\eqref{hamiltonian}. We diagonalize the interacting Hamiltonian
exactly to obtain the electron and spin densities. Since the
interacting system does contain very rich physics, we restrict the
discussion in the following to the case of a dot with two
electrons. To be concrete, we consider the case of an InAs dot here,
where the effective mass is $m^* = 0.042 m_e$, Land\'e factor
$g_L=-14$ and dielectric constant $\epsilon=14.6$. In this system it
appears to be experimentally feasible to change the ratio of the SOCs
$g_1/g_2$ over a wide range.

\begin{figure}[tbp]
\includegraphics[width=0.99\columnwidth]{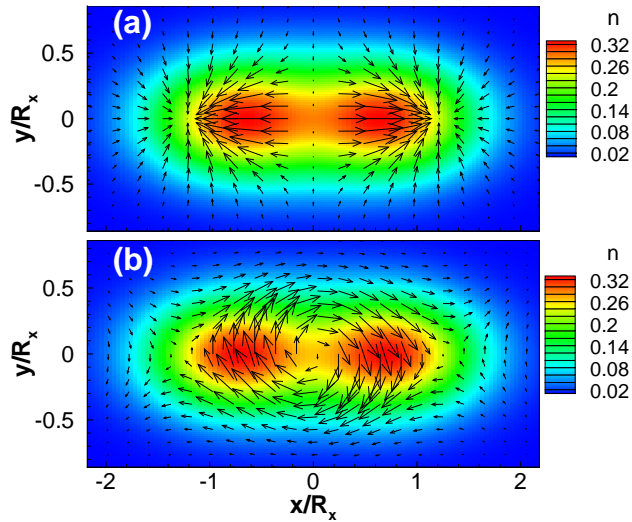}
\caption{(Color online) The in-plane spin fields in an elliptic dot with
two electrons, $R_x=15$nm, $R_y=10$nm at $B=5$T.  The colors represent
the electron density. (a) Rashba SOC only with $\hbar g_1=40$
nm$\cdot$ meV, and (b) Dresselhaus SOC only with $\hbar g_2=20$
nm$\cdot$ meV.}
\label{fig4}
\end{figure}

In a two-electron dot with Coulomb interactions, the spin textures can
be much more complex than in the single-electron case. If there is no
time reversal symmetry breaking, the texture is cancelled by the
Kramers pair. In the presence of a magnetic field, the spin textures
appear again with topological charge $+1$ or $-1$ if the dot is
perfectly isotropic. For an anisotropic quantum dot the electron
density will split into two centers in a strong magnetic field even
without SOC. With SOCs the spin textures are modified by this density
deformation. In the examples shown in Fig.~\ref{fig4}, we find in both
cases three vortices along the elongated $x$ axis.
In the Rashba SOC case shown in Fig.~\ref{fig4}(a) there are two
vortices with $q=1$ and one with $q=-1$, while there are two vortices
with $ q=-1$ and one with $q=1$ in the Dresselhaus SOC case presented
in Fig.~\ref{fig4}(b). Hence, the total winding numbers are still $+1$
and $-1$ in a Rashba SOC and Dresselhaus SOC system, respectively, as
in the single-electron dot. Indeed, the spin textures along the edges
of the dot are quite similar to the single-particle case. Here
interactions are less relevant and the spin textures are thus mainly
induced by the SOCs.

\begin{figure}[htp]
\includegraphics[width=0.99\columnwidth]{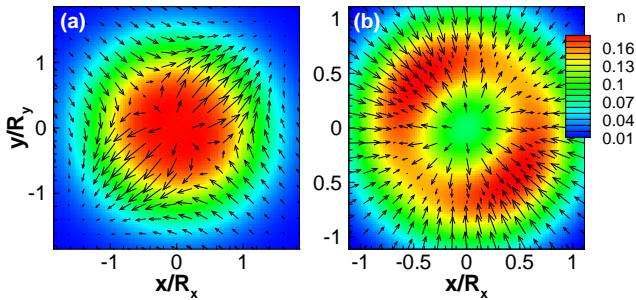}
\caption{(Color online) The in-plane spin fields in a two-electron dot with
$R_x=R_y=15$nm, and $\hbar g_1 =\hbar g_2 =20$ nm $\cdot$meV. The
colors represent the electron density. (a) $B=3.5$T, topological
charge $q=-1$, and (b) $B=18$T leading to $q=+1$.}
\label{fig5}
\end{figure}

In an isotropic two-electron dot with equal SOCs, $g_{1}=g_{2}$, we
find that both the density profiles and spin textures undergo a
dramatic change as a function of the applied magnetic field [Fig.~\ref{fig5}].
In this case, the spin and density profiles are determined
collectively by {\it both} the interactions and SOCs. For large
magnetic fields we find, in particular, that the electron density
splits mirror symmetrically along the line $x=y$ [Fig.~\ref{fig5}(b)],
causing also a complete rearrangement of the associated spin texture
and a change of the total topological charge. This has to be
contrasted with the case of an InAs dot without SOC where the angular
momentum of the ground state changes from $L=-1$ to $L=3$ at about
$B=17$T leading instead to a ring-shaped electron density. We further
note that in a ZnO dot with stronger Coulomb interaction \cite{aram},
the splitting of the electron density and the spin textures can be
generated in a much lower magnetic field. Details will be published
elsewhere. This splitting---which only occurs if both interactions and
SOCs are present---could possibly be observed experimentally and would
thus provide an indirect confirmation of a non-trivial spin texture in
the dot.


In summary, we find that the combination of electron confinement and
SOCs leads to vortex-like spin textures in the ground state even for a
single-electron dot. The spin texture can be stabilized by an external
magnetic field breaking the time-reversal symmetry. Interestingly, the
winding number of the vortex is different for dots with dominant
Rashba SOC or Dresselhaus SOC. This difference can be traced back to
the different symmetries of the Hamiltonian. The Rashba SOC commutes
with $L_z+\hbar \sigma_z/2$ leading to a topological charge of the
spin field of $q=+1$ while the Dresselhaus SOC commutes with
$L_z-\hbar \sigma_z/2$ and the topological charge is $q=-1$. Using
the exact diagonalization scheme we have shown that these spin vortices do
persist also in interacting multi-electron dots. For an elliptic
two-electron dot we find, in particular, that more than one spin
vortex can exist. In all investigated cases the total topological
charge is, however, still $q=\pm 1$ as in the single-electron
case. Physically, this is understood by noting that the spin
configuration at the edge of the dot, where the electron density is
low, is only weakly affected by the interactions. We thus conjecture
that the total topological charge for a spin texture in multi-electron
dots is always fixed to $q=\pm 1$.

The spin textures in QDs described in this letter are similar to the
in-plane structure of (anti-)skyrmion excitations in quantum Hall
systems. The locations of the skyrmions in a quantum Hall systems are,
however, unknown making it difficult to observe a single skyrmion
directly. The existence of skyrmions has so far only been confirmed
indirectly by NMR and transport measurements. In contrast, the spin
vortices in QD systems are localized at a known position. This might
possibly open new avenues for spintronics and quantum information
applications. Arrays of QDs have, for example, been realized
experimentally \cite{KouwenhovenHekking,PiqueroZulaica} and have been
considered as a potential platform for quantum computation
\cite{LossDiVincenzo,ZanardiRossi,Awschalom,Nowack}. In such an array of QDs with SOCs
the ratio of Rashba to Dresselhaus couplings might be tunable by gates
over a sufficiently wide range to realize a system with localized and
controllable topological charges $q=\pm 1$. At a minimum, such a setup
would allow for an indirect probe of the spin texture by measuring the
field dependence of the out-of-plane spin component [Fig.~\ref{fig3}],
either by a magnetometer or in an NMR experiment \cite{private}.

We acknowledge useful discussions with Sean Barrett (Yale). TC
acknowledges support by the Canada Research Chairs Program of the
Government of Canada. JS acknowledges support by the Natural Sciences
and Engineering Research Council (NSERC, Canada) and by the Deutsche
Forschungsgemeinschaft (DFG) via Research Unit FOR 2316. Computation
time was provided by Calcul Qu\'{e}bec and Compute Canada.

\clearpage
\newpage

\begin{widetext}
\huge{Unique Spin Vortices in Quantum Dots with Spin-orbit Couplings --
Supplemental material}\\*[0.3cm]
\large{Wenchen Luo, Amin Naseri, Jesko Sirker, and Tapash Chakraborty}\\

In the supplemental material we present details of the analytical
calculations for the single-electron parabolic quantum dot (QD) \cite{fock,darwin,QD_book}
and additional numerical data demonstrating the stability of the topological spin
textures described in the letter. In Sec.~1 the derivation of
the exact eigenstates for a single-electron dot with equal Rashba and
Dresselhaus spin-orbit couplings (SOCs) and without the magnetic field, as
well as the lifting of the Kramers degeneracy for small fields are
discussed. In Sec.~2 we present additional numerical data for
the single-electron QD showing that (i) for large magnetic fields and
$g_1=g_2$ the spin texture starts to rotate, (ii) the topological
charges are robust in the cases $g_1\gg g_2$ and $g_2\gg g_1$ and
similar to the spin textures with Rashba or Dresselhaus coupling only
discussed in the main text, and (iii) the topological charges are
also robust against straining of the QD. In Sec.~3 we give
explicit formulas for the Coulomb matrix elements for multi-electron
QDs.

\section{1. Single-electron dot: Equal Rashba and Dresselhaus couplings}
\label{Sec1}
The Hamiltonian of an electron in an isotropic dot
$\Omega_{x,y}=\Omega$, $\ell_{x,y}=\ell$ and $g_1=g_2=g$ without
magnetic field is equivalent to a two-component quantum Rabi model. In
this case the Hamiltonian (2) in the main text reads,%
\begin{equation}
H=\frac{\mathbf{p}^{2}}{2m^{\ast }}+\frac{1}{2}m^{\ast }\omega ^{2}\left(
x^{2}+y^{2}\right) -\left( p_{x}-p_{y}\right) g\left(\sigma_x + \sigma_y
\right) .
\end{equation}%
We now define the ladder operators of the quantum harmonic oscillator%
\begin{eqnarray}
a_{\mu } &=&\sqrt{\frac{m^{\ast }\omega}{2\hbar }}\left( \mu +\frac{i%
}{m^{\ast }\omega}p_{\mu }\right) ,
\end{eqnarray}%
and use $a_{x} =-ib_{x}, a_{y} =ib_{y},$ to transform the Hamiltonian to
\begin{eqnarray}
H =b_{x}^{\dag }b_{x}+b_{y}^{\dag }b_{y}+ G\left( b_{x}^{\dag
}+b_{x}+b_{y}^{\dag }+b_{y}\right) \frac{1}{\sqrt{2}}\left(\sigma_x + \sigma_y
\right) ,
\end{eqnarray}%
where $G= g\sqrt{\hbar m^{\ast }\omega}$. Then, performing  a
unitary transformation
\begin{eqnarray}
U &=&\frac{1}{\sqrt{2}}\left(
\begin{array}{cc}
1 & e^{-i \pi/4} \\
e^{i \pi/4} & -1%
\end{array}%
\right) ,
\end{eqnarray}
leads to
\begin{eqnarray}
UHU^{\dag } &=&b_{x}^{\dag }b_{x}+b_{y}^{\dag }b_{y}+ G\left(
b_{x}^{\dag }+b_{x}+b_{y}^{\dag }+b_{y}\right) \sigma_z ,
\end{eqnarray}%
which is a two-component quantum Rabi model with zero splitting \cite{rabi}.
The case of $g_1=-g_2$ can be solved in a similar manner.

In the main text, we show in Eq.~(5) that the ground states are a
degenerate Kramers pair. If the magnetic field is infinitesimal then the
degeneracy is lifted. Since ${}_{\pm}\left\langle GS\right\vert L_z
\left\vert GS\right\rangle_{\pm}=0$, the unique ground state can be found to
lowest order by minimizing the Zeeman energy only. We use the ansatz
\begin{equation}
\left\vert GS\right\rangle =A\left\vert GS\right\rangle _{+}+B\left\vert
GS\right\rangle _{-}.
\end{equation}%
with coefficients $A,B$. The Zeeman energy is then proportional to
\begin{eqnarray}
\left\langle GS\right\vert \Delta \sigma _{z}\left\vert GS\right\rangle
&=&-\Delta \int^{\infty}_{-\infty} dxdy\left( AB^{\ast }e^{-i2\sqrt{2}g
\ell \frac{m^*}{\hbar }\left( x-y\right) }+A^{\ast }Be^{i2\sqrt{2}g\ell
\frac{m^*}{\hbar }\left(x-y\right) }\right) e^{-x^{2}-y^{2}}  \notag \\
&=&-\Delta \left( AB^{\ast }+A^{\ast }B\right) \pi e^{-4\left( g\ell \frac{m^*
}{\hbar }\right) ^{2}}.
\end{eqnarray}%
For $\Delta >0$ the minimization of the Zeeman energy requires
$A=B=\frac{1}{\sqrt{2}}$ while it requires $A=-B=\frac{1}{\sqrt{2}}$
for $\Delta <0$. Hence,%
\begin{equation}
\left\vert GS\right\rangle =\frac{\left\vert GS\right\rangle _{+}+
\textrm{sgn}\left( \Delta \right) \left\vert GS\right\rangle _{-}}{\sqrt{2}}.
\end{equation}

The other calculations presented in the main text are the standard
first-order perturbative calculations.

\section{2. Single-electron dot: Numerical results}
\label{Sec2}
In Fig.~1(a) of the main text we have shown that the in-plane spin
texture in a weak magnetic field for the case $g_1=g_2$ is mirror
symmetric around the line $x=y$, consistent with the perturbative
calculation. For larger fields the perturbative ground state given above is,
however, no longer a good starting point. In Fig.~\ref{s1}, we show
how the in-plane spin textures evolve with increasing magnetic field.
\begin{figure}[tbp]
\centering
\includegraphics[width=12cm]{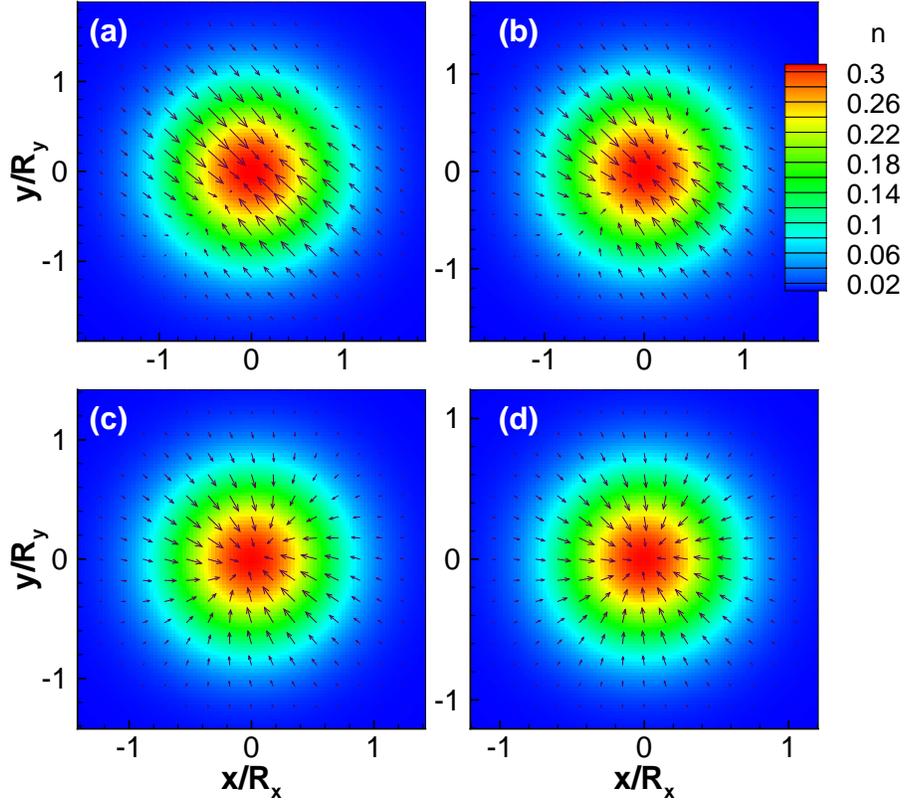}
\caption{(Color online) The in-plane spin field with $\hbar g_1= \hbar g_2=20$
nm$\cdot$ meV in a single-electron dot with $R_x=R_y=15$nm at magnetic
fields (a) $B=3$T, (b) $B=5$T, (c) $B=10$T, and (d) $B=15$T.}
\label{s1}
\end{figure}
All these results show a mirror symmetry about the line $x=\pm y$.  However,
when the magnetic field becomes stronger the spins start to rotate
leading to a spin texture similar to the case of Rashba SOC only. Note
that the in-plane spin components are weaker than in the Rashba case
though because the spin becomes more and more polarized along the
$z$-direction.

In Fig.~2 of the main text we have shown the spin vortices in a single-electron
dot if only the Rashba or the Dresselhaus SOC is present. In
Fig.~\ref{s2} we show that these results are indeed representative for
the regimes $g_1 \gg g_2$ and $g_2 \gg g_1$.
\begin{figure}[tbp]
\centering
\includegraphics[width=12cm]{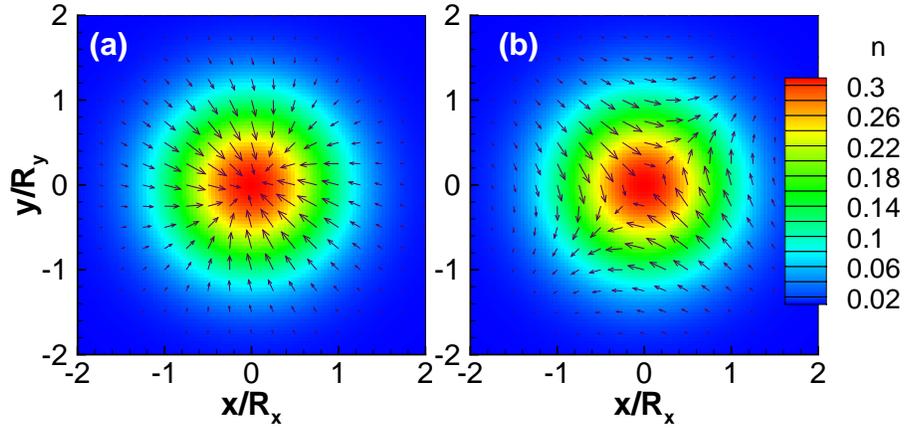}
\caption{(Color online) The in-plane spin field in an isotropic single-electron dot with
$R_x=R_y=15$nm at $B=0.1$T. The SOCs are (a) $\hbar g_1= 20$ nm$\cdot$
meV and $\hbar g_2=5$nm$\cdot$ meV, and (b) $\hbar g_1= 5$nm$\cdot$
meV and $\hbar g_2=20$nm$\cdot$meV.}
\label{s2}
\end{figure}
We also note that even for larger magnetic fields the topological
properties are not changed, although the spin textures are
weakened. States with higher topological charge $|q|>1$ may exist in the
excited states. Contrary to the spin textures in the ground state they
are, however, fragile due to their Kramers partner.

Finally, we also show that the spin texture is robust against
the eccentricity of the dot. We consider an elliptical InAs dot with
$R_x=15$nm and $R_y=10$nm. From Fig. \ref{s3} it is obvious that the
distortion of the dot does not qualitatively change the structure of
the spin vortex.
\begin{figure}[tbp]
\centering
\includegraphics[width=12cm]{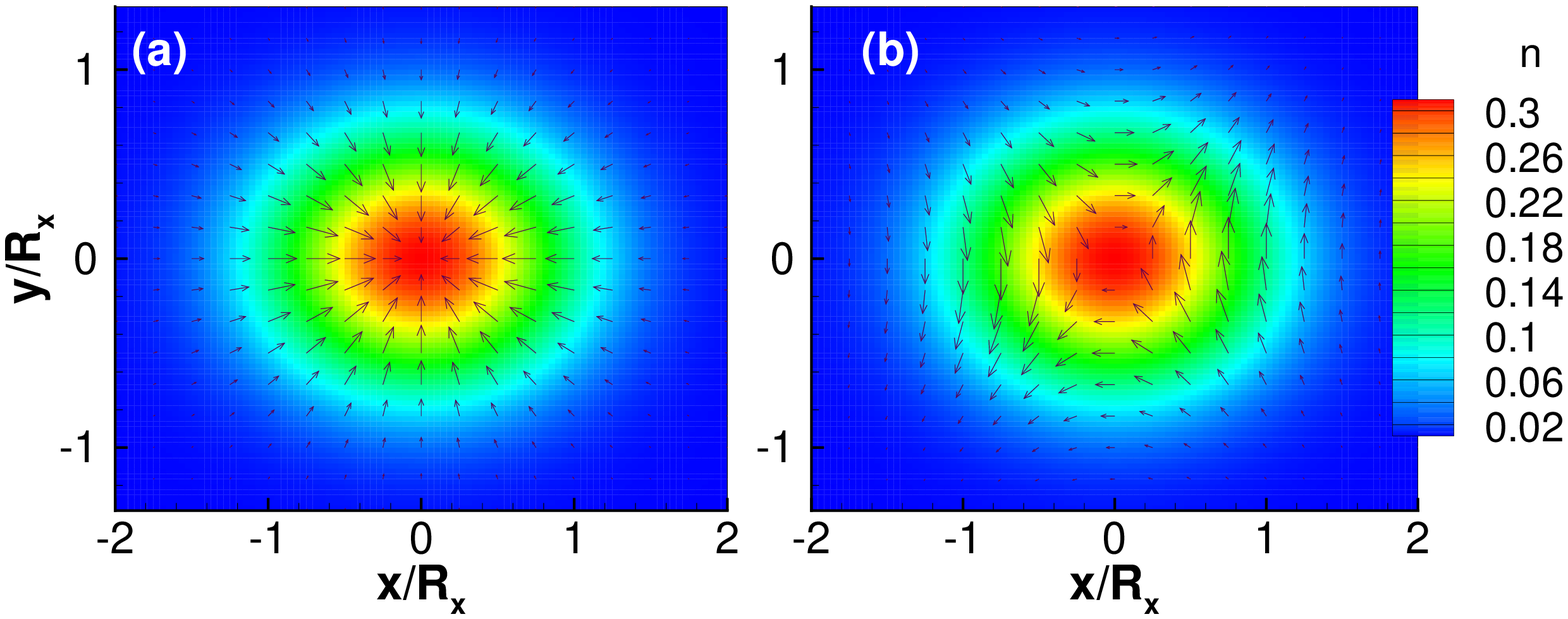}
\caption{(Color online) The in-plane spin field in a single-electron dot with $R_x=15$nm,
$R_y=10$nm in $B=0.1$T. The SOCs are (a) $\hbar g_1= 20$ nm$\cdot$ meV
and $\hbar g_2=0$, and (b) $\hbar g_1= 0$ and $\hbar
g_2=20$nm$\cdot$meV.}
\label{s3}
\end{figure}

\section{3. Multi-electron dots: Coulomb interaction matrix elements}
\label{Sec3}
Below we explicitly display the Coulomb interaction matrix
elements \cite{QD_book} used in the exact diagonalization method for multi-electron dots
\begin{eqnarray}
V\left( n_{1},n_{2},n_{3},n_{4}\right) &=&\frac{2}{\pi } \frac{e^{2}}{\epsilon \sqrt{\ell
_{x}\ell _{y}}}\phi \left( n_{1x},n_{4x}\right) \phi \left(
n_{1y},n_{4y}\right)\phi \left( n_{2x},n_{3x}\right) \phi \left( n_{2y},n_{3y}\right)   \\
&& \left( -1\right) ^{\left\vert n_{2x}-n_{3x}\right\vert +\left\vert
n_{2y}-n_{3y}\right\vert } i^{\left\vert n_{1x}-n_{4x}\right\vert +\left\vert
n_{1y}-n_{4y}\right\vert +\left\vert n_{2x}-n_{3x}\right\vert +\left\vert
n_{2y}-n_{3y}\right\vert }  \nonumber \\
&&\int_{0}^{\infty }dx dy\, \Phi \left( n_{1x},n_{4x},x\right) \Phi
\left( n_{2x},n_{3x},x\right)  \frac{\Phi \left( n_{1y},n_{4y},y\right)
\Phi \left( n_{2y},n_{3y},y\right) }{\sqrt{\frac{\ell _{y}}{\ell _{x}}x^{2}+\frac{%
\ell _{x}}{\ell _{y}}y^{2}}}. \nonumber
\end{eqnarray}%
Here $\epsilon$ is the dielectric constant and
\begin{eqnarray*}
\phi \left( n,m\right)  &=&\sqrt{\frac{2^{\min \left( n,m\right) }\min
\left( n,m\right) !}{2^{\max \left( n,m\right) }\max \left( n,m\right) !}},
\\
\Phi \left( n,m,x\right)  &=&x^{\left\vert n-m\right\vert }e^{-\frac{1}{4}%
x^{2}}L_{\min \left( n,m\right) }^{\left\vert n-m\right\vert }\left( \frac{%
x^{2}}{2}\right)
\end{eqnarray*}%
with the Laguerre polynomial $L$.

\end{widetext}

\end{document}